\documentclass[aps, prl, twocolumn, amsmath, amssymb, floatfix]{revtex4-1}

\usepackage[utf8]{inputenc} 
\usepackage[T1]{fontenc} 
\usepackage{graphicx} 
\usepackage{amsmath} 
\usepackage{amssymb} 
\usepackage{hyperref} 
\usepackage{siunitx} 
\usepackage{geometry} 
\usepackage{natbib}
\setcitestyle{square, comma, numbers,sort&compress, super}

\begin{document}

\title{Stochastic motility energetics reveals cooperative bacterial swarming in optical tweezers}

\author{Clara Luque-Rioja}
\affiliation{Translational Biophysics, Instituto de Investigación Sanitaria Hospital Doce de Octubre (Imas12), Madrid, Spain}
\affiliation{Departamento de Química Física, Universidad Complutense de Madrid, Madrid, Spain}

\author{Horacio López-Menéndez}
\affiliation{Departamento de Química Física, Universidad Complutense de Madrid, Madrid, Spain}
\affiliation{Escuela de Arquitectura y Tecnologia, Universidad San Jorge, Zaragoza, Spain.}

\author{Macarena Calero}
\affiliation{Translational Biophysics, Instituto de Investigación Sanitaria Hospital Doce de Octubre (Imas12), Madrid, Spain}
\affiliation{Departamento de Química Física, Universidad Complutense de Madrid, Madrid, Spain}

\author{Niccoló Caselli}
\affiliation{Translational Biophysics, Instituto de Investigación Sanitaria Hospital Doce de Octubre (Imas12),  Madrid, Spain}
\affiliation{Departamento de Química Física, Universidad Complutense de Madrid, Madrid, Spain}

\author{Diego Herráez-Aguilar}
\affiliation{Instituto de Investigaciones Biosanitarias, Universidad Francisco de Vitoria, Ctra. Pozuelo-Majadahonda, Pozuelo de Alarcón, Madrid, Spain}

\author{Juan P.G. Villaluenga*}
\affiliation{Departamento de Estructura de la Materia, Física Térmica y Electrónica, Universidad Complutense de Madrid, Madrid, Spain}

\author{Francisco Monroy*}
\email{monroy@ucm.es}
\affiliation{Translational Biophysics, Instituto de Investigación Sanitaria Hospital Doce de Octubre (Imas12), Madrid, Spain}
\affiliation{Departamento de Química Física, Universidad Complutense de Madrid, Madrid, Spain}

\renewcommand{\vec}[1]{\boldsymbol{#1}}
\newcommand{\tens}[1]{\boldsymbol{#1}}
\newcommand{\av}[1]{\left\langle #1 \right\rangle}
\newcommand{\abs}[1]{\left| #1 \right|}
\newcommand{\grad}{\vec{\nabla}}
\newcommand{\laplace}{\nabla^2}
\newcommand{\transpose}[1]{{ #1 }^{T}}
\newcommand{\diracdelta}[1]{ \delta\left( #1 \right) }
\newcommand{\RE}[1]{\text{Re}\left[#1\right]}
\newcommand{\IM}[1]{\text{Im}\left[#1\right]}
\newcommand{\tr}[1]{\mathrm{tr}\left[ #1 \right] }
\newcommand{\diag}[1]{\mathrm{diag}\left[ #1 \right] }
\renewcommand{\det}[1]{\mathrm{det}\left[ #1 \right] }
\newcommand{\inv}[2]{\mathcal{I}_{#1}^{#2} }
\newcommand{\identity}{\tens{1}}
\newcommand{\kbt}{k_\text{B}T}
\newcommand{\pea}{\text{Pe}_\text{a}}
\newcommand{\pes}{\text{Pe}_\text{s}}
\newcommand{\gp}{G'}
\newcommand{\gpp}{G''}

\newcommand{\ie}{\textit{i.e.} }
\newcommand{\eg}{\textit{e.g.} }
\newcommand{\etc}{\textit{etc.} }
\newcommand{\etal}{\textit{et.~al.}}
\newcommand{\apriori}{\textit{a priori} }

\newcommand{\fig}[1]{\textbf{Fig.~\ref{#1}}}
\newcommand{\FIG}[1]{\textbf{Figure~\ref{#1}}}
\newcommand{\eq}[1]{\textbf{Eq.~\ref{#1}}}
\newcommand{\EQ}[1]{\textbf{Equation~\ref{#1}}}
\newcommand{\sctn}[1]{\textbf{\S~\ref{#1}}}
\newcommand{\tbl}[1]{\textbf{Table~\ref{#1}}}
\newcommand{\movie}[1]{\textbf{movie~\ref{#1}}}
\newcommand{\movies}[1]{\textbf{movies~\ref{#1}}}


\definecolor{pumpkin}{rgb}{1.0,0.4,0.0}
\definecolor{mygreen}{rgb}{0.0,0.55,0.3}
\definecolor{strawberry}{rgb}{1.0,0.0,0.5}
\definecolor{midnight}{rgb}{0.003921569,0.098039216,0.576470588}
\definecolor{saphire}{rgb}{0.0,0.196,0.372549}
\definecolor{crimson}{rgb}{0.75686,0,0.262745}
\definecolor{capri}{rgb}{0.0,0.768627,0.8745098}

\newcommand{\jmr}[1]{{\color{blue}[JMR: {#1}]}}
\newcommand{\cv}[1]{{\color{cyan}[CV: {#1}]}}
\newcommand{\kt}[1]{{\color{mygreen}[KT: {#1}]}}
\newcommand{\ac}[1]{{\color{pumpkin}[AC: {#1}]}}
\newcommand{\tns}[1]{{\color{strawberry}[TNS: {#1}]}}
\newcommand{\tocite}[1]{{\color{cyan}XXX}}

\begin{abstract}
Bacterial flagellar swarming enables dense microbial populations to migrate collectively across surfaces, often resulting in emergent, coordinated behaviors. However, probing the underlying energetics of swarming at the single-cluster level remains a challenge. Here, we combine optical tweezers and multiparticle tracking within a stochastic thermodynamic framework to characterize the active motility of confined \textit{Proteus mirabilis} clusters. Using the Photon Momentum Method to directly measure trapping forces, we show that swarming clusters generate persistent, dissipative flows indicative of non-equilibrium stationary motility within confined solenoidal mesostructures. These flagellar rotational dynamics break detailed balance in mesoscopic force space and exceed the limits of passive friction, as evidenced by force–velocity correlations and vortex-like circulations. By coarse-graining cluster trajectories into an active Brownian phase space, we quantify the work performed by bacterial swarms at cooperative coupling to thermal fluctuations, resulting in dissipative Ohmic-like currents overcoming conservative trapping. Our findings establish a generalizable approach to quantify collective motility and energetic dissipation in active bacterial clusters, offering new insights into the physical principles governing microbial cooperativity.

\end{abstract}

\maketitle

In nature, non-equilibrium arises from persistent energy imbalances between systems and their surroundings \cite{boltzmann1872sitzungsberichte,chandrasekhar1943stochastic}. Power-consuming systems operating in fluctuating environments, such as living organisms, are thus constrained by fundamental limits on thermodynamic efficiency \cite{wang2006maximum}. Within the framework of stochastic energetics \cite{sekimoto1998}, these bounds remain valid even when systems exhibit Brownian-like diffusivity while maintaining non-equilibrium steady states (NESS) \cite{seifert2012stochastic,ciliberto2017experiments,peliti2021stochastic,bechinger2016active}. In NESS, thermal and active fluctuations coexist, persistently breaking detailed balance through correlated forcing that overcomes friction \cite{seifert2012stochastic, battle2016broken}. Despite their irreversibility, these mechanical transitions remain thermodynamically consistent at mesoscopic scales, in accordance with stochastic energetics \cite{sekimoto1998, egolf2000equilibrium}. NESS are characterized by continuous energy fluxes —as entropy production \cite{PRIGOGINE}, which quantifies the degree of irreversibility and governs heat dissipation through stochastic fluctuations \cite{bechinger2016active,landi2021irreversible,di2024variance}. Thermodynamic performance in such systems involves minimizing frictional dissipation while sustaining a positive entropy production rate \cite{PRIGOGINE,sekimoto2010stochastic, di2024variance}, consistent with adaptive responses observed in biological systems~\cite{dieball2024thermodynamic,li2019quantifying,aoki1995entropy,martin2001comparison,turlier2016equilibrium,lynn2021broken}. Microbial motility —including flagellar cell swimming, collective swarming and chemotactic migration, illustrates these principles, emerging as a canonical example of mesoscopic NESS constrained by fluctuating energetic landscapes~\cite{Purcell1977}. Indeed, the nonreciprocal in-milieu propulsion enables robust microbial adaptation to variable environments under thermal noise ~\cite{Purcell1977,lucia2015bioengineering,goel2016stochastic}. Hence, motion efficiency in microbial collectives depends on their thermodynamic consistency \cite{Purcell1977}, mediated by the coupling of metabolic power input, frictional resistance and resulting entropy production~\cite{cates2015motility,patteson2018propagation}. Bacterial cooperativity may then be arise as an optimized NESS, balancing dissipation with adaptivity under environmental randomness~\cite{skinner2021improved,deng2021measuring}.

Flagellated microbes are highly efficient at converting metabolic energy into swimming motion~\cite{schavemaker2022flagellar}. Their rotary motors --powered by chemical gradients under ATP consumption, are among the most complex molecular machines in biology~\cite{armitage2020assembly, manson1977protonmotive}, with evolved function adaptive to environmental responsiveness~\cite{tusk2018subunit}. While the biochemical basis of flagellar motility is well established~\cite{gottschalk1986regulation}, its energetic cost has mainly been inferred from indirect methods, including microcalorimetry, hydrodynamic analyses and computational modeling~\cite{cates2015motility, lauga2016bacterial, nirody2017biophysicist, kitao2018molecular, koch2011collective, elgeti2015physics}. Jones et al.~\cite{jones2021stochastic} used optical tweezers (OT) with the Photon Momentum Method (PMM)~\cite{bustamante2021optical, farre2010force, Farré2017} to measure mesoscopic forces in biflagellated algae \textit{Chlamydomonas reinhardtii}, revealing confined oscillations associated with its \textit{run-and-tumble} motility~\cite{polin2009chlamydomonas, tailleur2008statistical}. They quantified power dissipation up to $10^6 k_BT/\mathrm{s}$ ($\approx fW$), highlighting the energetic cost of breaststroke-driven propulsion. Battle et al.~\cite{battle2016broken} demonstrated broken detailed balance in \textit{Chlamydomonas}' flagellar beating by analyzing probability fluxes in coarse-grained phase spaces (CGPS)~\cite{sekimoto1998, seifert2012stochastic, sekimoto2010stochastic}, revealing mesoscopic NESS emerged upon fluctuation–dissipation theorem violations. Despite these advances, direct measurements of collective motility work during bacterial swarming remain unexplored, especially when multiple motors operate within clustered cells. While individual bacteria can be tracked \cite{MELL2018}, the limited number of trajectory samples complicate the thermodynamic analysis \cite{SALINAS2022}, making microscopic work inaccessible \cite{martin2001comparison, lynn2021broken}. Averaging over multiple single-cell trajectories is yet required for thermodynamic consistency \cite{sekimoto1998}, but this can obscure details of mesoscopic motion \cite{seifert2012stochastic, shinkai2014energetics}, a gap critical for understanding flagellar efficiency in bacterial clusters \cite{cates2015motility}.

The Photon-Momentum Method (PMM) enables direct measurement of ensemble-averaged forces without assuming a specific mesoscopic displacement model (e.g., a linear spring)~\cite{farre2010force}. Combining PMM-enabled optical tweezers (PMM-OT)~\cite{bustamante2021optical,zhang2019manipulating,farre2010force,Farré2017} with Multiple Particle Tracking (MPT)~\cite{soni2003single,rodriguez2015direct,herraez2020multiple} allows precise quantification of motility-induced work in trap-confined environments~\cite{speck2016stochastic,mandal2017entropy,goel2016stochastic}. PMM-OT/MPT evaluates absolute work along center-of-mass trajectories~\cite{jones2021stochastic}, assisted of displacement acquisition via imaging~\cite{MELL2018,SALINAS2022}. With sub-piconewton sensitivity and kilohertz sampling, it accurately resolves mechanical forces and dissipation during bacterial motility, offering insight into the thermodynamics of active systems. When integrated with a stochastic CGPS framework, this method quantifies flagellum-driven swarming under thermal noise without relying on linear force–displacement assumptions. Unlike traditional calibration-based approaches, PMM-OT/MPT captures real-time stochastic work fluctuations in non-equilibrium steady states (NESS), enabling direct assessment of active mechanical dissipation in bacterial clusters.

\subsection*{Motility setup: bacterial swarms} In this work, we focus on \textit{Proteus mirabilis}, a facultatively anaerobic, Gram-negative bacterium commonly found in soil and water \cite{armbruster2018pathogenesis}, here selected for its exceptional motility \cite{douglas1979measurement, matsuyama2000dynamic}. \textit{P. mirabilis} is an opportunistic pathogen causing wound, urinary, and septic infections \cite{armbruster2018pathogenesis}, often resistant to antibiotics \cite{chen2012proteus}. Its genus name, from the shape-shifting Greek god Proteus, reflects its adaptive ability to alter form and motility in response to environment \cite{armbruster2012merging}. Indeed, \textit{P. mirabilis} is an efficient swimmer in liquid media, also capable of collective migration on solid surfaces \cite{hoeniger1965development, Pearson2019}. In liquid environments, it exists as a vegetative rod-shaped bacillus (1 $\mu m$ long, 0.5 $\mu m$ width), equipped with 4 to 10 rotating flagella that enable swimming at a peak velocity, $c_S\approx 20 \, \mu m/s$ \cite{armbruster2018pathogenesis}. In gel cultures, broth viscosity suppresses free swimming \cite{tuson2013flagellum}, instead promoting a confined swarming state under surface confinement \cite{matsuyama2000dynamic}. On solid surfaces, \textit{P. mirabilis} elongates and produces more flagella than in its vegetative form \cite{armbruster2012merging}, reflecting ability to adapt to confining environments enhancing surface migration \cite{matsuyama2000dynamic, armbruster2012merging}. While flagellar mechanics constrains single-cell energetics~\cite{Purcell1977,nirody2017biophysicist}, collective bacterial motion involves hidden internal degrees of freedom~\cite{armitage2020assembly}. In \textit{P. mirabilis} swarms, thermodynamic consistency emerges from metabolically regulated adaptation to mechanical cues, ensuring efficient energy use~\cite{schavemaker2022flagellar,armbruster2012merging}.

\begin{figure} [htb]
    \centering
        \includegraphics[width=1.0\linewidth]{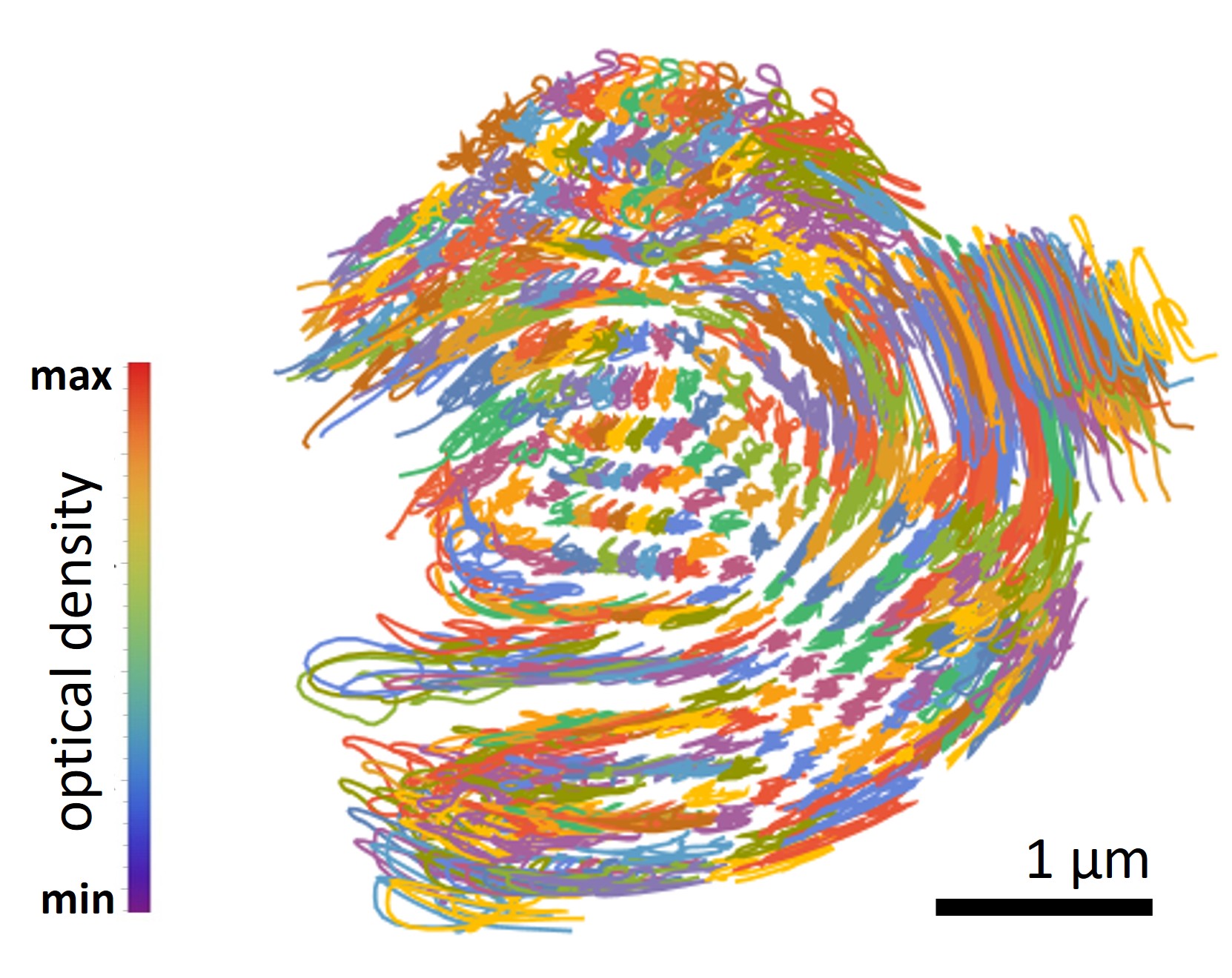}
    \caption{\textbf{Bacterial swarming motion detected via Multiple Particle Tracking (MPT) in a single optical tweezer.} Each detectable "particle" within the bacterial cluster is individually tracked to infer global center-of-mass motion (see Methods). Core-confined particles exhibit Brownian motion with collective rotational behavior, while peripheral, less constrained particles travel longer distances, following propelled trajectories originating from a central pivoting stroke. The color scale indicates particle masses inferred from optical densities (minimum relative to water density)}.
    \label{fig:tracking}
\end{figure}

As a proof of concept, we confined live \textit{P. mirabilis} swarms using optical tweezers and quantified their motility using PMM-OT force measurements~\cite{Farré2017} and high-resolution MPT~\cite{rodriguez2015direct,herraez2020multiple}. Trapped clusters exhibited coordinated, vortex-like motion under confinement —\textit{like a swarm within a hive} (Fig.~\ref{fig:tracking}, Suppl. Movie M1), consistent with active Brownian dynamics~\cite{patteson2018propagation}. From stationary NESS trajectories~\cite{gardiner1985handbook}, we computed motility work directly from center-of-mass displacements and optical forces~\cite{jones2021stochastic}. This stochastic energetics framework captures how internal dissipation sustains collective propulsion against trapping, revealing power–friction imbalances that constrain the thermodynamic efficiency of flagellar cooperativity in fluctuating microenvironments.

\begin{figure*}[htb]
    \centering
    \includegraphics[width=17.8cm,height=8cm]{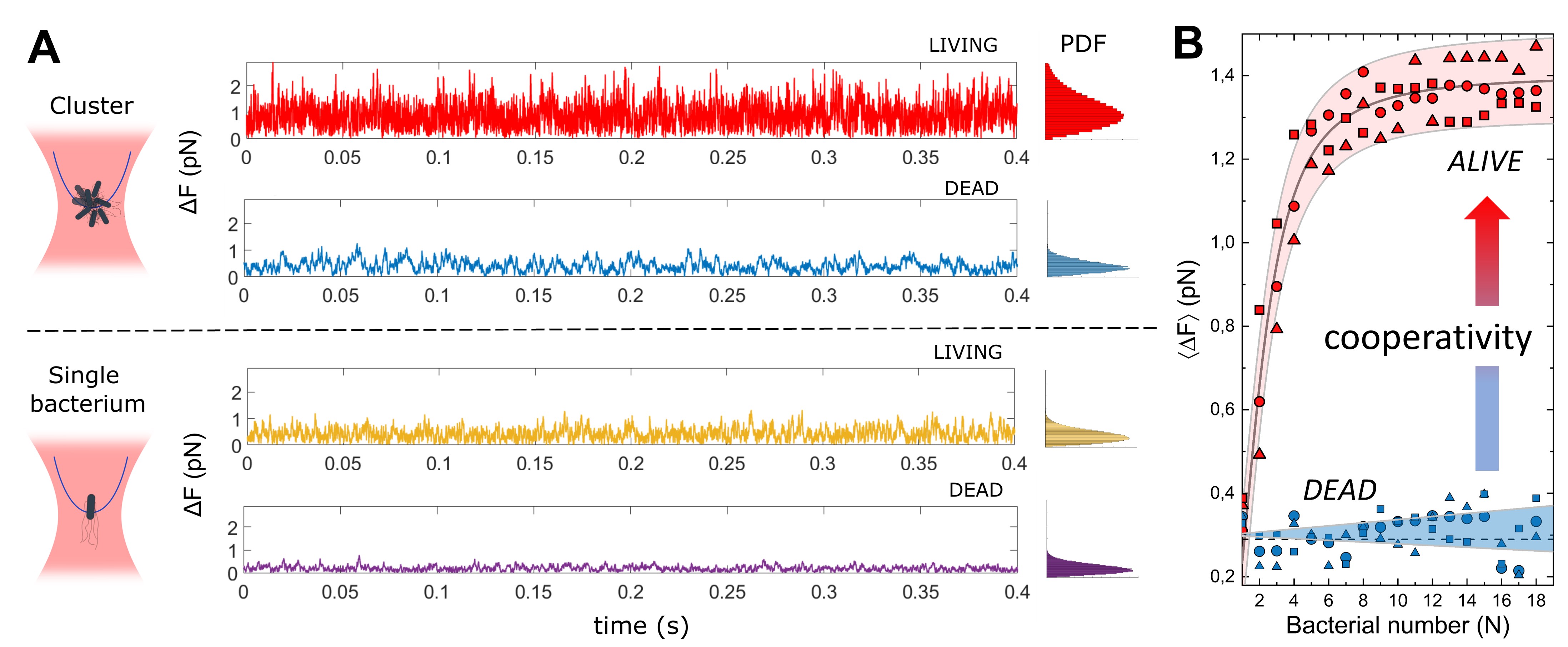}
    \caption{\textbf{Optical tweezers with photon-momentum method (OT-PMM) for trapping P. mirabilis bacteria.} Experimental conditions: laser power, $23 mW$ in sample; trapping stiffness, $k = 16-20 pN/\mu m$; solvent viscosity, $\eta = 10^{-3} kg/m \cdot s$; at room temperature, $T=22^{\circ}C$: \textbf{A) Fluctuating force realizations} for living and dead systems over time, $\delta F(t)$ (central panels): clusters (top); single bacteria (bottom). \textbf{Force probability distributions} follow Maxwellian modulus statistics, $P(F) \propto (F^2/\Delta F^2) exp(-F^2/2 \Delta F^2) $, with an effective force variance $\Delta F^2 \propto k_B T_{eff}$ (PDFs; right panels). \textbf{B) Realization-averaged fluctuating force} $\langle \Delta F (N)\rangle \equiv ({\bar \delta F^2})^{1/2} $ as a function of the number of bacteria aggregated into trapped clusters ($N$). Motile bacteria trapped alive (red); dead bacteria under fixation (blue). Symbols represent different aggregation sequences. Swarming cooperativity is observed as a logistic increase in motility forces, indicating functional binding among clustered living bacteria. The solid line represents the best fit to a Hill function (Eq. 1) with a critical cluster size of $N_S=9 \pm 2$, a Hill coefficient (inverse cooperativity) of $\alpha = 0.21 \pm 0.12$, and a saturation force of $\langle \Delta F \rangle_{\max} = 1.3 \pm 0.2$ pN. No force increase was detected in dead clusters relative to single bacteria. Dashed regions indicate variability bands.}
    \label{fig:LP_forces}
\end{figure*}

\section*{RESULTS}
We examined the confined motion of \textit{P. mirabilis} bacteria in culture broth using a single optical tweezer (OT) operated in Photon Momentum Method (PMM) mode to measure forces directly as ${\bf F} = d {\bf p}/dt$ \cite{ashkin1987optical, farre2010force}. Trapped cells exhibited in-plane diffusion under lateral forces, $\textbf{F}(x,y) \equiv (F_x,F_y) = \textbf{F}_0 + \delta \textbf{F}$, where $\textbf{F}_0$ represents the reaction force balancing the trap near the center-of-mass ($\textbf{F}_0 = -\textbf{F}_{trap}=k \bf{u}$; Suppl. Figs. S1A-B). As shown in Fig. \ref{fig:LP_forces}, both components align with the trapping $z$-beam’s axis, ensuring $F_z = 0$ at the focal plane ($z = 0$) \cite{Farré2017}. Bacterial clusters were assembled by capturing single cells with a “fishing” tweezer and transferring them into a sensing trap (Fig. \ref{fig:LP_forces}A, left panels), enabling direct-force measurements while minimizing photodamage (see Methods). Each cluster maintained a fixed number of viable, motile bacteria ($N$),  behaving as a cohesive, mobile entity (Fig. \ref{fig:tracking}; Suppl. Movie M1). Cluster sizes, $d(N)$, were visually assessed, ranging from single-cell size, $d_0 = 1.0 \pm 0.1 \mu m$ ($N =1$), up to $d_{max} \approx 4 \mu m$ for the largest clusters ($N \approx 20$) (Suppl. Fig. S1C). OT-PMM was performed under stable trapping conditions, with stiffness increasing linearly with cluster size, $k = 16 \rightarrow 20 \,pN/\mu m$ (Suppl. Figs. S1D-E). Experiments were initiated after confirming stationary reciprocal force trapping (${\bf F}_0 + {\bf F}_{trap} \Rightarrow $ zero offset) \cite{farre2010force}. Fluctuating components, $\delta\textbf{F}(N) \geq \textbf{F}_0$, increased with cluster size $N$, though the mean trapping force remained consistent across all clusters regardless of shape or size ($-\textbf{F}_{trap} \approx \textbf{F}_0 \approx 0.5 pN$; Suppl. Fig. S2). Relevant fluctuating forces were directly recorded using OT-PMM, as $\delta \textbf{F}_i \equiv (F_x, F_y)$, though they can be used to infer lateral displacements proportional to mobility impulses, $\Delta {\bf r}_i \equiv {\bf u}(x,y) \propto \delta \textbf{F}_i / k$ \cite{Farré2017}. Clusters were modeled as non-equilibrium steady-state (NESS) systems evolving in a coarse-grained phase space (CGPS), with active internal forces opposing external optical constraints \cite{CATES_ACTIVEFIELDS}. Control measurements using dead clusters and inert beads of similar size ($d \approx 1 - 4 \mu m$) confirmed that observed fluctuations were due to active motility (Suppl. Fig. S2).

\subsection*{Stochastic motility forces: from single bacteria to cooperative clusters}

Figure~\ref{fig:LP_forces} shows the stochastic force behavior of \textit{P. mirabilis} cells, either in clusters (top panels) or as single cells (bottom panels), including both living and dead states. Time-series of lateral forces $\delta \textbf{F}_i$ under stationary trapping reveal the variance $\Delta F^2 = \langle \delta \textbf{F}^2 \rangle = \langle F_x^2 + F_y^2 \rangle$, quantifying motility strength under confinement. Living clusters showed markedly higher $\Delta F$ compared to isolated cells, alive or dead. The force distributions $P[F(N)]$ depend on motility state: active clusters generate strong forces, $\Delta F_{act} \gg \Delta F_{pass} \approx F_0$, unlike dead cells, where $\Delta F_{pass} \approx k_BT/\delta \approx 0.2{-}0.3$~pN with $\delta \sim 10{-}20$~nm~\cite{armitage2020assembly, manson1977protonmotive}. These distributions follow Maxwell statistics, $P(\delta F) \propto \exp{[-\Delta W(\delta F)/2\Delta F^2]}$, where fluctuating work is $\Delta W = \delta F \Delta r \approx \delta F^2/k$ (Suppl. Fig. S3). Living clusters typically exhibited $\Delta F_{act} \approx 1.2{-}1.5$~pN, indicating non-equilibrium flagellar propulsion~\cite{cates2015motility}, while dead clusters remained near thermal equilibrium ($\Delta F_{pass} \approx \sqrt{k k_BT} \approx 0.2$~pN)~\cite{ashkin1987optical}. Figure~\ref{fig:LP_forces}B demonstrates that active forces increase with cluster size $N$, starting at $\Delta F_0 = 0.28 \pm 0.12$~pN ($N=1$) and saturating at $\Delta F_{act}^{(S)} \approx 1{-}1.5$~pN for $N \ge N_S \approx 10$. This cooperative amplification resembles allosteric chemotaxis~\cite{PARKINSON}, as modelled by the Hill function:

\begin{equation}
    \frac{\langle \Delta F \rangle}{\langle \Delta F\rangle _{max}} = \frac{N^\alpha}{N_S^\alpha+N^\alpha}.
\end{equation}

Here, $\alpha$ reflects initial steepness and $N_S$ marks the swarm cooperative onset (Fig.~\ref{fig:LP_forces}B). Swarming criticality arises as $\langle \Delta F \rangle_{max} \approx k_B T_{eff}(N_S)/\delta$, with $T_{eff}(N_S) > T$~\cite{cates2015motility}. In contrast, passive clusters exhibit no such scaling and remain limited to Brownian motion at $T_{pass} = T$ (Suppl. Fig. S3).

\begin{figure*} [htb]
    \centering
    \includegraphics[width=17cm]{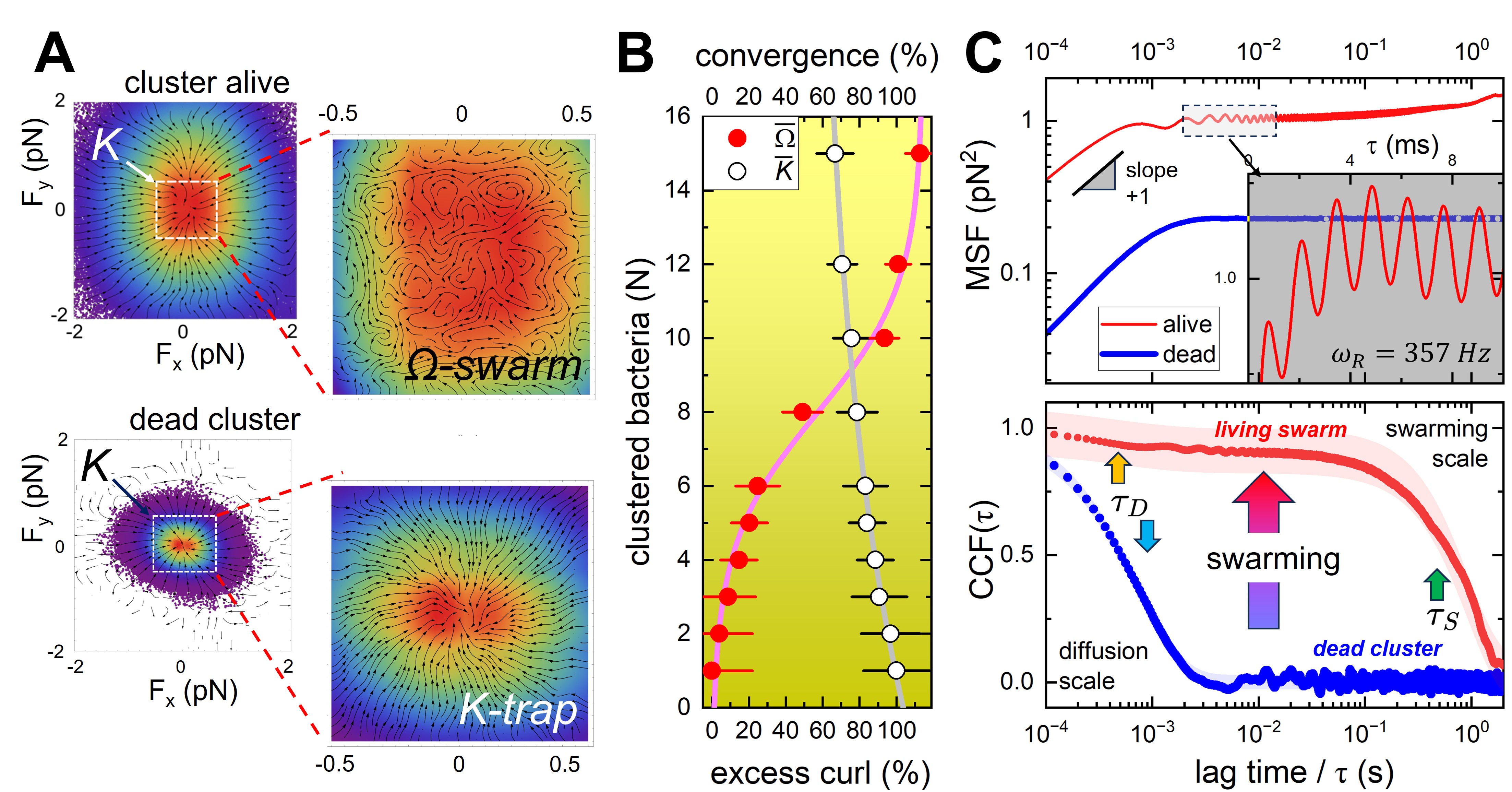}
    \caption{\textbf{CGPS fluctuating forces: A) Collective force currents} in living and dead clusters (left). Zoomed-in views show the most populated regions at low forces (right). In living clusters, swarming-broken detailed balance is seen as rotational $\Omega$-currents preventing trap attraction (top). In dead clusters, detailed balance appears as convergent currents towards central K-attractor (bottom). \textbf{B) Force-field relative characteristics} in living clusters: excess curl, $\bar J_\Omega \equiv (J_\Omega^{act}-J_\Omega^{pass})/J_\Omega^{act}$; relative convergence, $\bar J_K =J_K^{act}/J_K^{pass}$ (normalised means and bootstrapped standard deviations). \textbf{C) Force correlations over lag times:} mean-square force impulses show short-time diffusivity followed by long-time confinement (top); the rotatory flagellar mode appears at intermediate crossover times (inset). Normalized cross-correlation functions (bottom) reveal swarming activity at intermediate times ($\tau_D \le \tau \le \tau_S$), between rapid unconstrained diffusion ($\tau_D$) and terminal swarm confinement relaxation ($\tau_S$).}
    \label{fig:bdb}
\end{figure*}

\subsection*{Fluctuating force components: Coarse-grained phase-space decomposition}

To evaluate motility efficiency in bacterial clusters, we examined whether active force correlations establish a mechanically consistent swarming mechanism within a coarse-grained phase space (CGPS) \cite{cates2015motility}. By distinguishing non-thermal (correlated) from thermal (uncorrelated) fluctuations, we analyzed how dissipation in active swarms differs from that in inactive clusters diffusing in the optical tweezer’s deformation–work space. This dynamic space is defined by in-plane forces, ${\bf F}(x,y)$, driving overdamped displacements, ${\bf u}(x,y) \propto \delta {\bf F}/k$ (see Methods). Figure~\ref{fig:bdb} shows lateral fluctuating forces $\delta {\bf F} = (F_x, F_y)$, directly measured in CGPS under stationary probability, $dP(F_x,F_y)/dt = 0$ \cite{battle2016broken}. The resulting NESS force currents, ${\bf J}_F = P(\delta {\bf F}) \, d{\bf F}/dt$, describe stationary bacteromotive fluxes, linked to ensemble-averaged power injection, $\mathcal P = \int {\bf J}_F \cdot d{\bf u}$ \cite{jones2021stochastic, sekimoto1998}. In Fig.~\ref{fig:bdb}A, living clusters show scale-dependent force flow fields (top), contrasting with the purely overdamped motion of dead clusters (bottom). At large scales ($u \gg \delta$), trapping dominates ($F_{trap} \approx ku \approx 2$~pN), driving convergence to an equilibrium attractor ${\bf F}_K = 0$ at maximum probability $P_{max}$ (left panels). At microscopically zoomed scales ($u \approx \delta$), active swarms display vortex-like force fields ($\boldsymbol{\Omega}$), disrupting trapping convergence near ${\bf F}_K$ (Fig.~\ref{fig:bdb}A, top right). These vortices exceed local trapping forces ($\delta F_{act} \approx 0.5{-}1$~pN $\geq k\delta \approx 0.2$~pN) and generate spatial correlations consistent with NESS \cite{patteson2018propagation}. Even a single motile bacterium disrupts the convergent flow seen in passive beads (Suppl. Fig. S5). These observations support a decomposition into conservative and dissipative forces, ${\bf F} = {\bf F}_{cons} + {\bf F}_{diss}$, interpreted as momentum fluxes within a Navier–Stokes framework (NS), where the propagation speed of hydrodynamic interaction is defined as $c \equiv k/\eta$, set by the ratio of trap stiffness $k$ to internal viscosity $\eta$ (see Supplementary Note N1). 

\subsection*{Trapping convergence and active vorticity: broken detailed balance in bacterial swarms}

The optical tweezer imposes a conservative force field, ${\bf F}_{cons} = -k {\bf u} = -\nabla U$, derived from a scalar potential $U_{trap} \approx \frac{k}{2} {\bf u} \cdot {\bf u}$, where ${\bf u}$ represents CGPS deformation from center-of-mass displacements. Dissipative contributions arise from frictional and swarming forces: ${\bf F}_{diss} = {\bf F}_{frict} + {\bf F}_S = \nabla \times {\bf A}$, emerged from a generating vector potential, $\bf A = {\bf F}_{frict} \times \bf u$, capturing torque-like structured friction (Suppl. Note N1). The viscous friction force, ${\bf F}_{frict} = -\hat{\boldsymbol{\zeta}}(t) {\bf v}$, depends on the anisotropic friction tensor $\hat{\boldsymbol{\zeta}}$, where diagonal elements reflect hydrostatic resistance ($\zeta_{ii} \propto a\eta$), and off-diagonal terms encode directional coupling ($\zeta_{ij} \ne 0$). This rank-2 tensor may vary in time depending on the active swarming forces, $\hat{\boldsymbol{\zeta}}[{\bf F}_S(t)]$, with the curling tensor defined as $\hat{\boldsymbol{\kappa}} \equiv \partial \hat{\boldsymbol{\zeta}}/\partial t$. A negative determinant indicates lubrication i.e., $Det[{\boldsymbol{\hat \kappa}}({\bf F}_S)]<0$, reflecting frictional modulations driven by the active swarming force, ${\bf F}_{S} = + \hat{\boldsymbol{\kappa}} \space \bf u$ (Suppl. Note N1). 

The Helmholtz–Hodge (HH) decomposition separates the CGPS force field into gradient and rotational components: $\bf{F} = \bf{F}_G + \bf{F}_R$, with ${\bf F_G} = -\nabla U_{trap} - \frac{1}{c} \partial_t \bf A$ and $\bf{F}_R = \nabla \times \bf A$. Here, $\partial_t \bf A = {\bf F}_S \times \bf v$ represents frictional dissipation, and $c = k/\eta$ is the propagation speed. These field components underpin the NS–dynamics governing swarming:

\textbf{A. Convergent flows} result from trap deformation and frictional pressure gradients:
\begin{equation}
K = -\nabla \cdot {\bf F} = k + \nabla^2 p - \nabla \cdot {\bf F}_S \Rightarrow -\nabla \cdot \mathbf{F}_G,
\label{CONVERGENCE}
\end{equation}
with ${\bf F}_{trap} = -k {\bf u}$, and ${\bf F}_S^{\parallel} = +\boldsymbol{\kappa}_{ii} {\bf u}$. Conservation of the convergent flow, $\nabla \cdot {\bf F}_{S}=0$, requests on the stringent condition $\nabla\cdot {\bf A}=0$ (Coulomb gauge), constraining traceless curling, $\boldsymbol{{\kappa}}_{ii}=0$.

\textbf{B. Vortical currents} under incompressibility ($\nabla \cdot \bf v = 0$):
\begin{equation}
\boldsymbol{\Omega} = \nabla^2 (\hat{\zeta} \boldsymbol{\omega}) + \nabla \times {\bf F}_S \Rightarrow \nabla \times \mathbf{F}_R,
\label{VORTICITY}
\end{equation}
where $\boldsymbol{\omega} = \nabla \times \bf v$ is the angular velocity. Transverse swarming forces ${\bf F}^\perp_S = +\hat{\boldsymbol{\kappa}}_{ij} {\bf u}$ generate vorticity $\boldsymbol{\Omega}$ through anisotropic modulation. Coupling terms include $\nabla \times {\bf F}_S = \text{Tr}(\hat{\boldsymbol{\kappa}}) \nabla \times {\bf u} + \text{Antisymm}(\hat{\boldsymbol{\kappa}}) \nabla \cdot {\bf u}$, linking hydrostatic and deviatoric curling components (for details, see Suppl. Note N1).

Therefore, the dynamics of swarming are governed by $\hat{\boldsymbol{\zeta}}({\bf F}_S; \eta)$ and its derivative $\hat{\boldsymbol{\kappa}}({\bf F}_S; \eta)$, forming an adaptive NS-balance between convergence and lubricated circulation (Fig.~\ref{fig:bdb}A). This structure defines NESS energy flows ${\bf J}(E) = d{\bf F}[E(k,\zeta,\kappa)]/dt$, resolved via HH–decomposition: ${\bf \bar J}(E) = {\bf \bar J}_K + {\bf \bar J}_\Omega$. The force-field components are integrated over CGPS (Suppl. Fig. S6): $\bar K = \Vert \nabla \cdot {\bf F} \Vert \Rightarrow \nabla\cdot \bf F_G$ (Eq.~\ref{CONVERGENCE}), and $\bar \Omega = \Vert \nabla \times {\bf F} \Vert \Rightarrow \nabla \times {\bf F}_R$ (Eq.~\ref{VORTICITY}). Figure~\ref{fig:bdb}B shows $\bar K \approx 95{-}100\%$ and $\bar \Omega \approx 0$ for $N=1$, transitioning to $\bar \Omega \approx 100\%$ and $\bar K \approx 50\%$ for $N \geq N_S \approx 9{-}10$. This swarming onset marks a shift in energy transformation: from dominant convergence (${\bf \bar J}_K \gg {\bf \bar J}_\Omega$) under structural friction (${ \bf F}_{frict} \gg {\bf F}_S$), to dominant rotation (${\bf \bar J}_\Omega \gg {\bf \bar J}_K$) under active forces (${\bf F}_S \gg {\bf F}_{frict}$). As a rule of thumb, the addition of Dextran—a metabolically inert thickener—caused frictional currents to increase with viscosity (Suppl. Fig. S7). This confirms that detailed balance is broken under active frictional dissipation, where the net dissipative force ${\bf F}_{\mathrm{diss}} \,(= {\bf F}_{\mathrm{frict}} + {\bf F}_S) \gg {\bf F}_{\mathrm{cons}}$ dominates \cite{battle2016broken}. Swarm inactivation via glutaraldehyde restored overdamped balance (${\bf \bar J}_\Omega = 0$; Fig.~\ref{fig:bdb}A, bottom). Hence, we deduce correlated swarming forces driving lubricated vorticity, enhancing energy dissipation under broken frictional symmetry.

\begin{figure*}
    \centering
    \includegraphics[width=\textwidth]{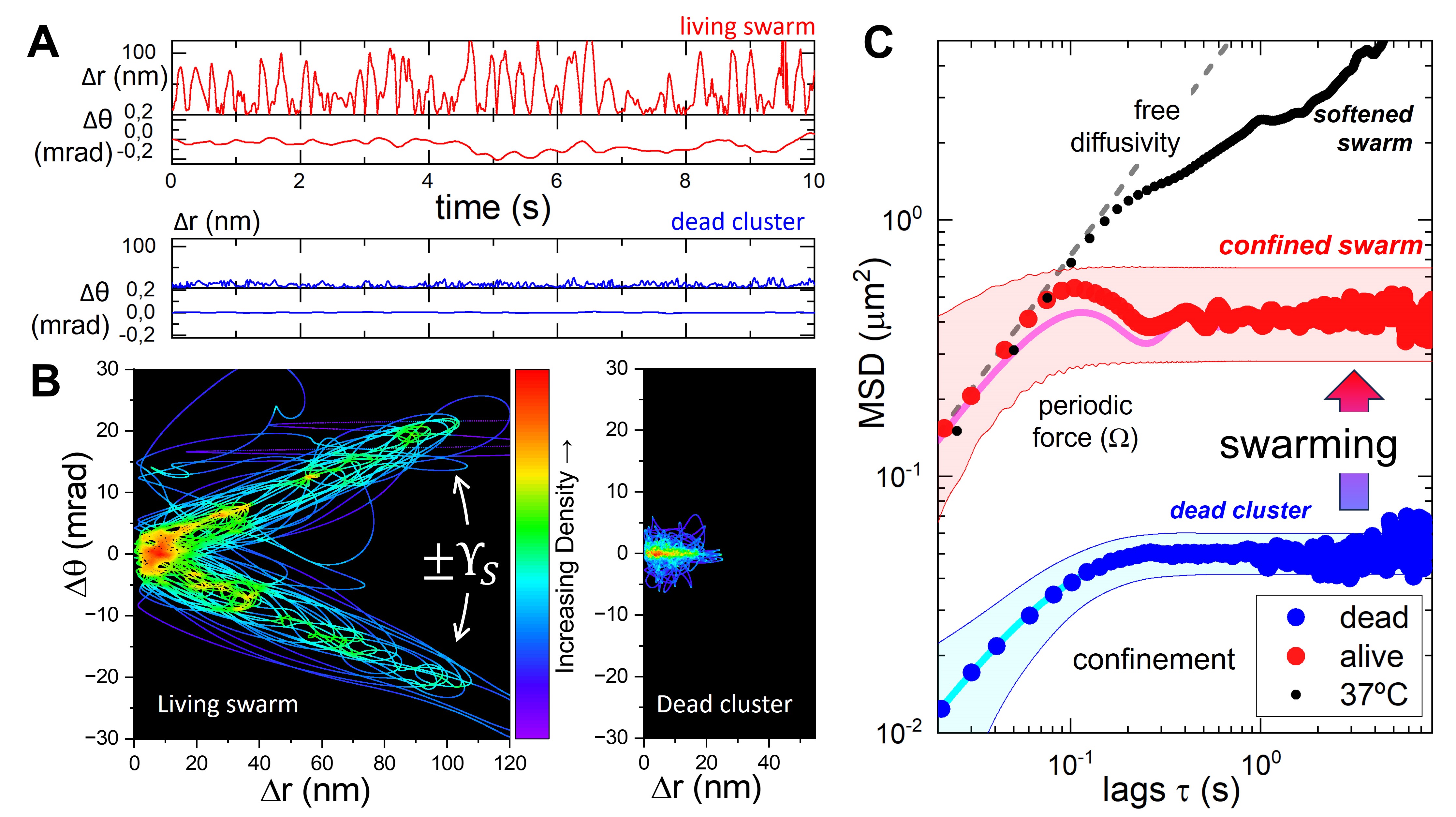}
    \caption{\textbf{Real-space cluster displacements in optical traps: A) Patterns in polar displacements over time} (distribution histograms; Suppl. Fig. S7). \textbf{B) Centre-of-mass' positions over time} showing directed movements in  a bacterial swarm maintained alive. The dead cluster undergo confined Brownian motion. \textbf{C) Mean square displacements:} Dead (blue) and living clusters (red) at room temperature ($22\, \text{ºC}$). The dashed regions represent variability bands. Straight lines represents best fittings to confined diffusivity processes: overdamped Ornstein-Uhlenbeck in dead clusters (cyan); underdamped Brownian oscillator in living clusters (magenta); Einstenian free-diffusivity (grey); see main text for details. At physiological temperature ($37\, \text{ºC}$), living swarms develop longer displacements as an effective trap-softened hyperdiffusivity (black symbols).}  
    \label{fig:displacements}
\end{figure*}

\subsection*{Force-to-force time correlations: fast diffusion and slow swarming}

We analyzed force time-series to identify causal correlations in bacterial clusters. Figure \ref{fig:bdb}C shows the mean-squared force fluctuations, defined as $MSF(\tau) \equiv \langle [\textbf{F}(t+\tau) - \textbf{F}(t)]^2 \rangle$, over different lag times for trapping mobility ($\tau$). At short timescales ($\tau \leq \tau_D \approx 1$ ms), the $MSF$ grows linearly, $MSF_{diff}(\tau) \approx \Delta F_0^2 (\tau/\tau_D)$, driven by thermal impulses $\Delta F_0 \approx k_B T/\delta$, with $\delta \approx 10$ nm. Beyond this ($\tau \gg \tau_D$), a plateau appears: $MSF_{conf} \approx \Delta F_{eff}^2 (\tau/\tau_D)^0$, representing confined fluctuations, where $\Delta F_{eff}^2 \approx (k_B T_{eff}) k$, and $T_{eff} \ge T$ for $k \approx 20$ pN$/\mu$m \cite{seifert2012stochastic, battle2016broken}. Living clusters exhibited forces over an order of magnitude greater than dead cells, indicating active propulsion ($\Delta F^2_{act} \gg \Delta F^2_{pass}$), or equivalently, $T_{act} \approx 10T$. Notably, oscillatory features emerged near crossover times ($\tau_{cross} \approx 10 \tau_D$), revealing flagellar beating at $\omega_R \approx 357$ Hz \cite{manson1977protonmotive, armitage2020assembly}. Force relaxation was further characterized by the normalized cross-correlation function, $\overline{CCF}(F_x,F_y; \tau)\equiv [\langle F_x(t)F_x(t+\tau) \rangle + \langle F_y(t)F_y(t+\tau)\rangle - 2\langle F_x(t) F_y(t+\tau) \rangle]/2\Delta F^2$ (Fig. \ref{fig:bdb}D). For inactivated cells, this correlation decayed as a single exponential, $\overline{CCF}_{pass}(\tau) \sim \exp(-\tau/\tau_D)$ with $\tau_D = 0.73 \pm 0.04$ ms, consistent with thermal motion ($D_0\equiv \delta^2 /2 \tau_0$), where the displacement amplitude is estimated as $\delta \approx 10 nm (\Leftarrow k_B T/F_0)$. At this microscopic scale, diffusivity aligns with the passive elastohydrodynamic response speed, estimated as $c = k/\eta \approx 400 \,\mu\text{m/s} \, (\approx \delta/\tau_0)$, assuming an effective internal cluster viscosity of $\eta \approx 50 \,\text{mPa} \cdot \text{s}$ \cite{tuson2013flagellum}. In contrast, living clusters exhibited two relaxation timescales: a fast component ($\tau_D = 0.3 \pm 0.2$ ms, $A \simeq 10\%$) and a dominant slow swarming mode ($\tau_S = 0.6 \pm 0.2$ s, $90\%$ amplitude). Thus, CGPS force-correlations reveal a dynamic transition from rapid Brownian motion to slower coordinated swarming.

\subsection*{In-trap bacterial displacements by multiple particle tracking: cooperative curling} 

Kinematic differences between living and dead bacterial clusters were assessed via real-space trajectories in optical traps (Fig. \ref{fig:tracking}). Fast videomicroscopy with Multiple Particle Tracking (MPT) \cite{herraez2020multiple} allowed non-invasive estimation of velocities as ${\bf \dot r}(t) = \Delta \textbf{r}/\Delta t$ from optical density variations (Suppl. Movie M1). Center-of-mass trajectories, recorded as planar paths ${\bf r}(t) = (x,y)$, appeared highly directional in swarming clusters but random in dead ones (Suppl. Fig. S8). Real-space displacements, expressed in polar coordinates ${\bf \Delta}_i = (\Delta r_i, \Delta \theta_i)$ relative to the trap center, were independently measured. Figure \ref{fig:displacements}A displays time-series of radial displacements showing oscillatory active motion in swarms (upper panels) versus passive fluctuations in dead clusters (lower panels). Living swarms oscillated at sub-second frequencies (3–4 cycles/s; $\omega_S \approx 20 \, s^{-1}$) and exhibited preferred angular displacements, yielding non-Gaussian orientation distributions unlike the symmetric Gaussians in dead clusters (Suppl. Fig. S9). These oscillations matched the slow collective relaxation time in CGPS force correlations ($\tau_S$), with $\omega_S \gtrsim 2\pi/\tau_S$ (Fig. \ref{fig:bdb}D). The much slower $\omega_S$ compared to intrinsic flagellar rotation ($\omega_R \approx 10^3 \,s^{-1}$; Fig. \ref{fig:bdb}C, inset) suggests mesoscopic emergent behaviour. This NESS-like motility reflects coupling between swarming forces and confinement relaxation \cite{patteson2018propagation}. Figure \ref{fig:displacements}B shows polar displacement trajectories ${\bf \Delta}(t) = [\Delta r(t), \Delta \theta(t)]$ exhibiting alternating pivot-like oscillations with angular gain $\Upsilon_S = \dot{\theta}/\dot{r} \approx 0.2 \, mrad/nm$, consistent with a confined \textit{“tug-of-war”} between swarming directions ($\pm\Upsilon_S$). This antisymmetric motion implies rotational structure, contributing to CGPS vorticity (see Supplementary Note N2). Thus, swarms maintain nonzero vorticity (\ref{fig:displacements}A, left panel), unlike isotropic, non-vortical dead clusters (right panel).

\subsection*{Diffusive mean-square displacements (MSDs): swarming oscillatory correlations}

We analyzed mean-squared displacements (MSD) computed over lag times as $MSD(\tau) = \langle [\mathbf{r}(t+\tau)-\mathbf{r}(t)]^2 \rangle$ (see Methods). Figure~\ref{fig:displacements}C compares living swarms with dead clusters, both showing confined diffusivity consistent with an Ornstein–Uhlenbeck process in 2D, as $MSD(\tau) \approx 4 D(\tau)\,\tau_{\rm conf}\left(1 - e^{-\tau/\tau_{\rm conf}}\right)$. The apparent diffusion coefficient is given by $D(\tau) = D_{\rm eff}^{(0)} + \Delta D_S(\tau)$, where $D_{\rm eff}^{(0)} = k_B T_{\rm eff}/\zeta_0$, represents an effective Einsteinian term (with $\zeta_0$ as isotropic solvent friction), and $\Delta D_S(\tau)$ accounts for active swarming contributions \cite{gardiner1985handbook, battle2016broken}. The confinement time, $\tau_{conf} = \zeta_0 / k$, depends on trapping stiffness ($k$), and Stokes friction by the external solvent ($\zeta_0 \approx 6\pi d \eta_0$, where $d$ is cluster size and $\eta_0$ is water viscosity). We observed $\tau_{\rm conf} \approx (0.5\!-\!1 )\,s$ (Fig.~\ref{fig:displacements}C), consistent with $k\approx20\,\mathrm pN/\mu\mathrm m$ and $\zeta_0\approx2\times10^{-5}\,\mathrm{kg/s}$~\cite{tuson2013flagellum}. At short times ($\tau \ll \tau_{\rm conf}$), $MSD\approx4D^{(0)}_{\rm eff}\tau$, with bare diffusion $D_0 = k_BT/\zeta_0 \approx 0.05\,\mu\mathrm m^2\!/\!s$. Dead clusters showed $D^{(0)}_{\rm pass} = (0.044 \pm 0.015)\,\mu\mathrm m^2\!/\!s$, while living swarms exhibited $D^{(0)}_{\rm act} = 2.2 \pm 0.2\,\mu\mathrm m^2\!/\!s$, indicating $T_{\rm eff}\gg T$. At $\tau \gg \tau_{\rm conf}$, MSD plateaued at $4\sigma^2$, with $\sigma^2_{\rm pass} \approx 0.048\,\mu\mathrm m^2$ vs. $\sigma^2_{\rm act} = 0.45 \pm 0.12\,\mu\mathrm m^2$, giving $T_{\rm eff}/T \approx 9 \pm 2$ and characteristic energy scale $\Delta E_{\rm act}\approx10k_BT$. Active swarming introduced an underdamped Brownian oscillator:  
\begin{equation}
 \Delta D_S(\tau) \approx \frac{\Delta E_{\rm act}}{\zeta}\sin^2(\omega_S\tau),
\end{equation} 

with $\omega_S \approx 20\,\mathrm s^{-1}$ and diffusive strength $\sigma_S^2 \omega_S \approx 0.5\,\mu\mathrm m^2\!/\!s$. The resulting swarming speed, $v_S = \omega_S/\Upsilon_S \approx 100\,\mu\mathrm m/\mathrm s$, remains slower than underlying elastohydrodynamic interaction, $c = 400\,\mu\mathrm m/\mathrm s$. This active oscillatory process occurred over mesoscopic distances ($\sigma_S \approx 150 \, nm \gg \delta$), and significantly longer curling times than merely diffusive flights ($\tau_S \ge \omega_S^{-1} \approx 50 ms> \tau_D \gg \tau_0$) (Fig. \ref{fig:tracking}). Notably, heating the swarm bath to physiological temperatures (37°C) reduced confinement constraints, resulting in longer swarming relaxation times ($\tau_S \ge \tau_{conf}$), and relaxed trapping distances ($\sigma_{act} > \sigma_S \gg \sigma_{pass}$) compared to room temperature (22°C) (Fig. \ref{fig:displacements}C, upper; softened swarm case). This metabolically induced relaxation increased radial displacements, allowing near-free diffusivity well beyond thermal limits, $D_{eff}^{(0)} \gg D_0 \equiv k_B T/\zeta_0 $ and $\sigma_{eff} \ge \sigma_S \gg \sigma_0 \equiv k_B T/k$. Metabolic swarm activity intensified in glucose-enriched media, further disrupting detailed balance, indicated by changing convergence and vorticity in CGPS (Suppl. Fig. S10).

\begin{figure*}[htb]
  \centering
    \includegraphics[width=17.8cm]{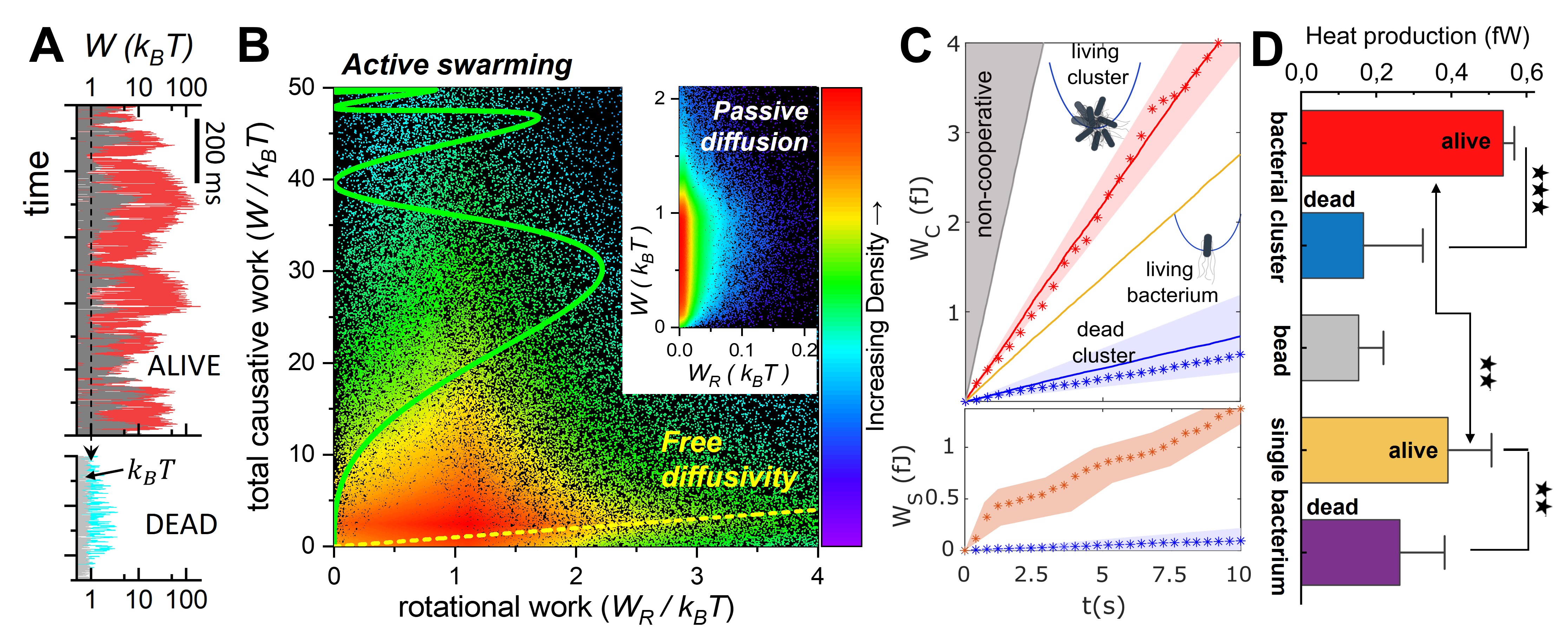}
    \caption{\textbf{Swarming energetics} measured from P. mirabilis clusters' center-of-mass motion. A) Experimental time series of work increments in bacterial clusters: live swarms (top) and dead clusters (bottom). Bursts of causative motility work ($W$, red) emerge with the presence of rotational energy $W_R$ (grey). B) Energy CGPS for motility work: Probability currents flow from high rotational work ($W_R$), driving increased total motility work depending on confined active diffusivity after lag time, $W[W_R,D(\tau)]$. Diffusive trajectories in parametric time: active swarming $W_R = \zeta\, \Delta E_{act}\, sin^2(\omega_S \tau)$ and $W = \zeta \,[D^{0}+\Delta D_S(\tau; \Delta E_{act}, \omega_S)]$ (green line: $\Delta E_{act}=20 k_BT$ and $\omega_S=20 s^{-1}$); passive free diffusivity $W = W_R=\zeta\, D^{(0)}$ (yellow dashed line). C) Cumulative work as a function of integration time (Top): path-independent (dots) vs. path-dependent (lines), for live clusters (red) and dead clusters (blue). Straight lines represents best fits to Ohm-like currents. (Bottom) Rotational work emerges as the swarming force imposes an Ohmic linear current; same symbols apply. D) Motility power, as rate of change of total mechanical energy, calculated from $W-t$ slopes in panel C (see main text for details).}
    \label{fig:work}
\end{figure*}

\subsection*{Swarming bursts: active diffusion space}

We characterized energy currents by quantifying motility work along trajectories: $d\bar W(\mathbf{F}, d\mathbf{r}) = \bar{\mathbf{F}}_i \cdot (\mathbf{r}_{i+1} - \mathbf{r}_i)$, using discrete Stratonovich multiplication, where $\bar{\mathbf{F}}_i = \{F_x, F_y\}$ denotes mean fluctuating force during $\Delta t = [t_i, t_{i+1}]$, and $\mathbf{r}_i = \{x, y\}$ is the center-of-mass position~\cite{sekimoto1998, sekimoto2010stochastic}. Figure~\ref{fig:work}A shows active clusters produce bursts lasting $\tau_S \approx (200 - 500) \,ms$ with work amplitudes from a few to several tens of $k_B T$, exceeding equilibrium energy $\langle d\bar W_{eq} \rangle \approx k_B T$. Rotational work was similarly defined as $d\bar W_R = \bar{\mathbf{T}}_i \cdot \Delta \theta_i$, with torque $\bar{\mathbf{T}}_i = \bar{\mathbf{F}}_i \times \mathbf{r}_i = (0, 0, \bar{T}_z)$, and $\bar{T}_z = \bar{F}_x \Delta y - \bar{F}_y \Delta x$. Active swarms exhibited significant rotational work ($d\bar W_R \approx k_B T$), while dead clusters showed negligible rotation (Fig.~\ref{fig:work}A, bottom). As shown in Fig.~\ref{fig:work}B, translational bursts were driven by swarming forces ($\Delta E_{act} \approx 10 k_B T$), with a secondary torque component $d\bar W_R \approx \Delta E_{act} \Upsilon_S (d/2) \approx k_B (T_{\rm eff}-T)$ reflecting vorticity (Suppl. Note N2). Passive clusters showed neither (Fig.~\ref{fig:work}B, inset). These observations align with a structured Navier–Stokes framework (Suppl. Note N1), where NESS swarming emerges from: \textit{(i)} translational divergence: $W - W_{\rm trap} \approx \hat{\zeta}_{ii}(t) D_{\rm eff} \approx k_B T_{\rm eff} \Rightarrow \nabla \cdot F_S \approx K - k$; and \textit{(ii)} rotational vorticity: $W_R \approx \hat{\zeta}_{ij}(t)\Delta D_S \approx k_B (T_{\rm eff} - T) \Rightarrow \nabla \times F_S \propto \Omega$ (see Eqs. 2-3). These act under thermal fluctuations satisfying $\zeta_0 D^{(0)} \approx k_B T \propto \nabla^2 p$~\cite{shinkai2014energetics}. Together, isotropic divergent bursts and anisotropic vortical curling support the oscillatory flagellar \textit{tug-of-war} (Figs.~\ref{fig:bdb},~\ref{fig:displacements}), governed by structured time-dependent friction $\hat{\boldsymbol{\zeta}}(t)$ and lubricated active forces $\mathbf{F}_S(\hat{\boldsymbol{\kappa}})=+\hat{\boldsymbol{\kappa}} \,\bf u$, contrasting with passive isotropic diffusion ($\hat{\boldsymbol{\kappa}}_{\rm pass} \equiv 0$). Notably, most swarming events occur within a single oscillatory half-cycle, above the activation threshold for unconfined motion (Fig.~\ref{fig:work}B).

\subsection*{Cumulative trapping work: frictional gauges}

To quantify energy flow in confined bacterial clusters, we analyzed center-of-mass trajectories $\Gamma[{\bf r}_i(t_i)] = (x_i, y_i)$ and calculated cumulative motility work using the Stratonovich integral~\cite{sekimoto1998}:
\begin{equation}
\bar{W}[\Gamma(t)] = \int_0^t d\bar W(t) \approx \sum_i \bar{\bf F}_i \cdot \Delta {\bf r}_i,
\end{equation}
where $\bar{\bf F}_i$ is the average fluctuating force over $\Delta t = [t_i, t_{i+1}]$, and $\Delta {\bf r}_i$ is the discrete displacement.

Alternatively, using the Hookean relation from the optical trap, ${\bf F}_{trap} = -k \Delta {\bf r}$, we inferred displacements from forces, as ${\bf dr}_i \Leftarrow {\bf F}_i/k$ \cite{farre2010force}. This framework yields a path-independent proxy for cumulative
work: 
\begin{equation}
\bar{W}_{\rm inf}(t) \approx \frac{1}{k} \sum_i \left(F_{x_i}^2 + F_{y_i}^2\right).
\end{equation}
Figure~\ref{fig:work}C shows ensemble averages ($N \approx 20$ cells per cluster, $>50$ realizations) confirming $\bar{W} \approx \bar{W}_{\rm inf}\, (=\bar W_{cons}-\bar W_{diss})$ within experimental uncertainty for both active and passive systems. As ensemble-averages, $\bar{W}_{cons}$ denotes the conservative work done by the trap on the system, while $\bar{W}_{diss}$ represents the dissipative work performed by the system against it \cite{sekimoto2010stochastic,seifert2012stochastic}. Energy transport is governed by trapping confinement (${\bf F}_{\rm trap} = -k\,{\bf u}$), viscous drag (${\bf F}_{\rm frict} = -\hat{\boldsymbol{\zeta}}\,{\bf v}$), and swarming propulsion (${\bf F}_S = +\hat{\boldsymbol{\kappa}}\,{\bf u}$), where ${\bf v} = d{\bf u}/dt$ is the dissipative velocity induced by trapping work against friction (see Supplementary Note N3). The\ energy loss rate is captured by the Rayleigh dissipation function~\cite{LANDAU}:
\begin{equation}
\Phi({\bf F}_{\rm diss}) = \frac{dW_{\rm diss}}{dt} = \frac{1}{2}\int {\bf F}_{\rm diss} \cdot d{\bf v},\quad {\bf F}_{\rm diss} = {\bf F}_{\rm frict} + {\bf F}_S,
\end{equation}

where the ensemble-averaged power reads:
\begin{equation}
\mathcal P = \langle\frac{dW}{dt}\rangle = \bar{K}_{\rm cons} - \bar{\Phi}_{\rm diss},
\end{equation}
as energy exchange between conservative trapping, $\bar{K}_{\rm cons} = c\,\langle {\bf F}_{\rm trap} \rangle$, upon elastohydrodynamic interaction at rate $c = k/\eta$, and total dissipation, $\bar \Phi_{diss}=\frac{1}{2}\langle {\bf F}_{diss}\cdot {\bf v}\rangle=\frac{1}{2}\langle {\bf F}_{frict}\cdot {\bf v} \rangle +\frac{1}{2} \langle {(\bf F}_{S}\times {\bf v})\cdot \bf z\rangle$ (Suppl. Note N3). In passive clusters, the trap generates converging flows that are opposed solely by isotropic friction, $\hat{\boldsymbol{\zeta}}_{\rm pass} = \zeta_{\rm iso} \mathbf{I}$ and $\bar{\Phi}_{\rm diss} \approx \bar{\Phi}_{\rm frict}^{\rm (pass)}$, yielding an equilibrium trade-off governed by in-trap elastohydrodynamics (Fig. \ref{fig:work}C, dead cluster; blue symbols). Hence, passive energy transfer obeys detailed balance, with conservative work offset by frictional losses: $\mathcal P_{pass}=\bar{K}_{cons} - \bar{\Phi}_{frict}^{(pass)}$ \cite{LANDAU}. By contrast, active clusters exhibit time-dependent deviatoric friction:
$ \hat{\boldsymbol{\zeta}}_{\rm act}(t) \approx \hat{\boldsymbol{\zeta}}_{\rm pass} - \hat{\boldsymbol{\kappa}}\,t$, where the swarming gauge, $\partial_t {\bf A} = {\bf F}_S \times {\bf v}=\hat{\boldsymbol{\kappa}}{\bf u}\times {\bf v}$, enters via lubricated curling $\hat{\boldsymbol{\kappa}}$. Thus, the active power budget becomes $\mathcal P_{\rm act} = \bar{K}_{\rm cons} - \bar{\Phi}_{\rm frict}^{\rm (act)} + \bar{\Phi}_S$, exceeding $\langle P_{\rm pass} \rangle$ due to swarming gains (Fig.~\ref{fig:work}C). Hence, energy balance reads:
\begin{equation}
\mathcal P_{\rm act} = \mathcal P_{\rm pass} - \bar{\Phi}_L(\bar{\kappa}) + \bar{\Phi}_S(\bar{\zeta}),
\end{equation}
where $\bar{\Phi}_L \equiv \bar{\Phi}_{frict}^{(pass)} - \bar{\Phi}_{frict}^{(act)} \propto  \bar\kappa \, \bar v^2$ is the lubrication gauge quantified as active energy input from assembled curling, $\bar \kappa= \langle \int \kappa_{ij}(t') dt'\rangle$ (antisymmetric tug-of-war), and $\bar{\Phi}_S = \langle{\bf F}_S \cdot d\mathbf{v} \rangle \propto R \,\bar v^2$ is the swarming gauge via frictional resistance, $R=\langle det (\zeta_{ij}) \rangle$ (see Supplementary Note N4).

Despite strong internal force fluxes in living swarms~\cite{koch2011collective}, motility power remains steady, $\mathcal P_{\rm act} \approx \bar J_\Omega^2 R$, indicating Ohmic-like dissipation governed by a constant frictional resistance ($R$). Hence, the average normal gauge potential, $\langle \partial_t {\bf A}_z \rangle = \langle {\bf F}_S \times {\bf v} \rangle_z = \bar{\Phi}_S - \bar{\Phi}_L =\mathcal P_{act} - \mathcal P_{pass}> 0$, reflects net energy flow sustained by deviatoric friction. Since $\bar{\Phi}_S$ arises solely from anisotropic (curling) components, it excludes contributions from convergent flows in $\bar{\Phi}_L$, which couples to isotropic friction. Thus, $\bar{\Phi}_S$ and $\bar{\Phi}_L$ together define the anisotropic energy balance under trapping (Suppl. Note N4).
\subsection*{Dissipated motility work: Ohmic currents}

Analogous to Joule heating, the mechanical power dissipated by swarming activity can be expressed via an Ohmic relation \cite{CATES_ACTIVEFIELDS,LANDAU}:
\begin{equation}
\mathcal P_{\rm act} = R_\zeta\, \bar{J}_\Omega^2,
\end{equation}
where \( R_\zeta = \langle \hat{\zeta}(\eta, {\bf F}_{\rm diss}) \rangle \) is the effective friction coefficient, and \( \bar{J}_\Omega \) denotes the coarse-grained rotational (vortical) dissipation current associated with swarming-induced curl (see Supplementary Note N5). The convergent component \( \bar{J}_K \) does not contribute to swarming energetics under this formulation, being absorbed by passive frictional dissipation. This Ohmic relation obeys the First Joule's law (Fig. \ref{fig:work}C):
\begin{equation}
\bar{W}_{\rm diss}(t) = \bar{\Phi}_{\rm diss} \, t \approx R_\zeta\, \bar{J}_\Omega^2 \, t,
\end{equation}
with effective resistance \( R_\zeta \approx \bar{V}_S/\bar{J}_\Omega \), where \( \bar{V}_S(\bar{F}_S, \bar{K}_{\rm cons}) \) is the mean swarming potential set by optical confinement. 

This frictional resistance scales linearly with the environmental viscosity (\( R_\zeta \propto \eta \)), as confirmed experimentally by adding Dextran to the medium (Suppl. Fig. S7). Since Dextran increases \(\eta\) without targeting the intrinsic motility force \( F_S \), this experiment isolates the impact of viscous damping on energy dissipation (see Suppl. Fig. S11A). Assumed constant \( \bar{V}_S \), the cumulative trapping work decreases with viscosity (\( \bar{K}_{\rm cons} \propto k/\eta \); S11B), while the dissipated power increases linearly with inverse resistance (\( \bar{\Phi}_{\rm diss} = \bar{V}_S^2/R_\zeta\); S11C). Hence, the NS-dissipation rate defines differential Ohmic gauges \( \bar{\Phi}_{\rm diss} = \bar{\Phi}_L - \bar{\Phi}_S \), reflecting how CGPS dissipative forces drive net vortical dissipation as $\bar{J}_\Omega = \bar{V}_S /R_\zeta \Leftarrow \langle {\bf F}_{\rm diss}(\zeta,\kappa) \times {\bf v} \rangle_{\Omega} > 0$. As a reference, the motility work performed by a single bacterium \( \bar{W}_0^{(1)} \) lies between passive (\( \bar{J}_\Omega = 0 \)) and collective swarming (\( \bar{J}_\Omega \gg 0 \)) (see Fig. \ref{fig:work}C). The assumed uncooperative bound \( \bar{W}_{\rm UC} = N \bar{W}_0^{(1)} \gg \bar{W} \) emphasizes the efficiency of swarm-level force coordination. The observed path-independence of \( \bar{W} \) under metabolism-fixed \( \bar{V}_S \) confirms a steady Ohmic regime, with power regulated by the interplay of frictional (\( \bar{\Phi}_L \)) and active (\( \bar{\Phi}_S \)) dissipation.

\subsection*{Swarming work: dissipative vortices}

Swarming-induced rotational work is quantified by the cumulative angular energy,
\begin{equation}
\bar{W}_R(t) = \int_0^t \mathbf{F}_S \cdot d\boldsymbol{\theta} \approx \sum_i (\mathbf{F}_i \times \mathbf{dr}_i) \cdot \hat{\mathbf{z}},
\end{equation}
tracking the work performed by active forces along angular displacements in the trapping plane. \\
Figure~\ref{fig:work}C (bottom) shows that only living swarms generate significant rotational work, scaling as $\bar{W}_R \propto \bar{J}_\Omega^2\, \bar{\kappa}\, t$, while dead clusters remain inactive ($\bar J_\Omega^{(\text{pass})} = 0$). The associated dissipation, $ \bar{\Phi}_\Omega = \frac{1}{2} \langle \mathbf{F}_{\rm diss} \times \mathbf{v} \rangle_z$, arises from the interplay between dissipative forces—primarily the swarming term $\mathbf{F}_S = + \hat{\boldsymbol{\kappa}} \mathbf{u}$, and rotational velocity fields from structured friction, $\mathbf{F}_{frict} = - \hat{\boldsymbol{\zeta}} \mathbf{v}$. The Ohmic character of this power stroke, $\bar{\Phi}_\Omega \propto \bar{\Phi}_S - \bar{\Phi}_L$, links active lubrication to net energy loss through confined vortical motion (Suppl. Note N4). In living \textit{P. mirabilis} swarms, active angular displacements scale with vorticity magnitude upon frictional resistance:
\begin{equation}
 \bar{W}_R \equiv \langle {\bf F}_S \times {\bf u} \rangle_z \approx T(\Omega)\Upsilon_S\, \Delta r \propto \frac{\Omega^2}{2R_\zeta},
\end{equation}

where the normal torque $T(\Omega) = (\nabla \times \mathbf{F}_S)_z$ captures the swarming-induced circulation. The observed rotational work confirms swarming vortices as a dissipative Ohmic-like current driven by internal friction under active lubrication (Suppl. Fig. S12). Moreover, increased rotational work correlates with enhanced diffusivity, $\bar{W}[\bar{W}_R, \Delta D_S(\omega_S)]$ (Fig.~\ref{fig:displacements}), indicating cooperative flagellar dynamics. These observations support a threshold of coordinated dissipation at critical swarm size $N \geq N_S$, where mesoscale vortices emerge as collective energy states (Fig.~\ref{fig:LP_forces}; \ref{fig:work}B). Similar mechanisms were reported in \textit{Chlamydomonas}, where beat-aligned rotation enhances oscillatory power extraction \cite{LEPTOS2023}.

\subsection*{Stochastic energetics: heat and entropy production}

The dissipated power of bacterial swarming can be analyzed via the stochastic energetics framework by Shinkai and Togashi \cite{shinkai2014energetics}, which relates effective diffusion to heat dissipation under non-equilibrium conditions \cite{chandrasekhar1943stochastic}. In our swarm system, the effective diffusion coefficient is decomposed as \( D_{\text{eff}}(\tau) = D^{(0)} + \Delta D_S(\tau) \), where \( D^{(0)} \) represents thermal fluctuations at effective temperature \( T_{\text{eff}} \), and \( \Delta D_S(\tau) = (\Delta E_S/\zeta)\, \sin^2(\omega_S \tau) \) is the time-dependent active component thermodynamically consistent with underdamped swarming (Fig.~\ref{fig:displacements}). During the active diffusion window (\( \tau \le \omega_S^{-1} \)), the cumulative motility work \( \bar{W}(\tau) = \int_0^{\tau} \mathbf{F} \cdot \mathbf{v} \, dt \) is dissipated by viscous forces \( \mathbf{F}_{\text{frict}} = -\boldsymbol{\hat{\zeta}} \mathbf{v} \) (Fig.~\ref{fig:work}C). The corresponding heat dissipation,
$ -\bar{Q}(\tau) = \bar{W}(\tau) - 2k_B T$, accounts for frictional losses minus the thermal baseline \cite{shinkai2014energetics}. Under energy conservation, \( \Delta E = \bar{W}_{\text{frict}} + \bar{Q} = 0 \), thus making the dissipation rate, \( \Delta \mathcal P = \mathcal P_{act} -  \mathcal P_{pass} = d\bar{W}_{\text{frict}}/dt = -d\bar{Q}/dt \), a direct measure of irreversible energetic cost \cite{jones2021stochastic}. This power imbalance enables estimation of the entropy production rate \( \sigma \equiv \Delta \mathcal P/T \ge 0 \), offering a thermodynamically consistent account of active dissipation \( \Delta \mathcal P = \bar{\Phi}_S - \bar{\Phi}_L \), and entropy generation in swarming vortical flows. Figure~\ref{fig:work}D shows gauges of dissipated power across conditions. Living swarms reach \( \mathcal P_S  = (1.3 \pm 0.2) \times 10^5 \, k_B T/\text{s} \),  consistent with active diffusion estimates \( \mathcal P_S \approx \Delta E_S/\tau_D \), yielding \( \sigma \approx 10^5 \, k_BT/\text{s} \). Dead clusters dissipate much less, \( \mathcal P_{\text{dead}}  \approx 3.2 \times 10^3 \, k_B T/\text{s} \), reflecting passive Brownian motion under trap confinement. Hence, optimal cooperative regulation via swarming-induced lubrication is gauged by a differential power of \(\Delta \mathcal{P}_S = (\bar{\Phi}_S - \bar{\Phi}_L)_{\text{opt}} \approx 10^5\,k_BT\,(\approx 0.5 \, fW)\). In comparison, single cells produce a similar power \( \mathcal P_1 \approx 0.3 \, fW \), whereas uncoordinated aggregates dissipate substantially more: \( \mathcal P_{\text{UC}} \equiv N \mathcal P_1 \approx 3 \, fW \gg \mathcal P_S \). In words, cooperative swarming realizes optimal mesoscopic dissipation well below the uncooperative bound, underscoring its optimal efficiency in non-equilibrium steady states (NESS) under frictional cooperativity.

\section*{DISCUSSION}
Our findings provide a mechanistic framework for understanding energy dissipation in bacterial swarms as non-equilibrium steady states (NESS) driven by internally structured friction and collective motility forces. By combining optical confinement with coarse-grained phase space (CGPS) analysis, we resolve the interplay between conservative trapping forces and  active, time-correlated fluctuations in living \textit{P. mirabilis} clusters under flagellar beating. Decomposing CGPS force-fields into gradient and solenoidal components interpreted on the prism of an active Navier-Stokes (NS) field reveals two distinct energetic modes underlying swarming cooperativity: conservative flows dominating passive response, and dissipative vortical currents arising from frictional self-interaction during active lubricated curling due to flagella power stroke. These NS-modes are encoded in the effective friction tensor $\boldsymbol{\hat \zeta}(t)$, which includes a time-independent isotropic baseline set by viscosity ($\zeta_{ii} \propto \eta$) and a time-dependent deviatoric part reflecting swarming-induced lubrication stresses ($\hat \zeta_{ij}$). The active component, quantified via the curling tensor $\boldsymbol{\hat \kappa} \equiv \partial_t \boldsymbol{\hat \zeta}$, sustains circulation and drives irreversible dissipation. We show that motility work follows a linear Ohmic-like scaling, $\bar W = \bar \zeta(\eta) \, v^2 \, t$, where the resistance determinant $\bar \zeta\Leftrightarrow R_\zeta$ captures structured friction in the bacterial ensemble. A rotational current, $\bar W_R = \bar \kappa \, \Omega^2 \, t$, emerges from dissipative vortices, highlighting the role of vortical structures mediated by antisymmetric $\boldsymbol{\hat\kappa}$, the time-dependent adaptive lubrication allowing for a rotational \textit{tug-of-war} around the swarm's centre-of-mass. Despite the complexity of underlying trajectories, the observed path independence at the coarse-grained level justifies ensemble-based energetic estimates. Swarming clusters exhibit motility power up to 40-fold higher than passive assemblies, yet remain below the uncoordinated active limit, indicating efficient cooperative transport. Thermodynamic analysis under the Shinkai–Togashi formulation confirms that motility-induced work balances dissipated heat, $-\bar W(\tau) \leftrightarrow \bar Q(\tau)$, with strictly positive entropy production. This supports our interpretation of bacterial swarms as adaptive NESS systems that optimize dissipation through spatiotemporal modulation of friction in a rotational \textit{tug-of-war} cooperative motion. The onset of rotational energy transport near the cooperative swarming threshold ($N \approx N_S$) marks a symmetry-breaking transition in force balance—from isotropic convergent relaxation to curl-driven lubricated circulation, providing a robust biophysical signature of collective activation.

In summary, our findings identify frictional curling self-interaction through the curling lubrication gauge as a key control parameter for swarming cooperativity, linking deformation, force transmission and energetic efficiency in confined bacterial collectives. The mechanical (Navier-Stokes) emergence of a rotational Ohmic current reveals a linearly regulated, metabolically sustained dissipative pathway driven by a simple vortical structure, reinforced by lubricated friction. This feedback mechanism enables swarms to exploit internal vortices as controlled agents of active transport, stabilizing collective motion near a reduced yet optimized cooperative NESS. By sustaining a rotational imbalance, the system thermodynamically enhances directional persistence while minimizing excess energetic cost—a robust strategy for adaptive motility under confinement. Our stochastic energetics framework, uniting theory and experiment, offers testable predictions for motility regulation and the design of synthetic swarms, where dissipation is precisely modulated by collective mechanics under confinement constraints.

\section*{CONCLUSION AND OUTLOOK}
This study establishes a quantitative mechanical link between bacterial swarming and non-equilibrium thermodynamics, using stochastic energetics to characterize energy flows in confined \textit{Proteus mirabilis} clusters. By integrating optical tweezers and multiple-particle tracking, we directly measured motility work and heat dissipation, revealing that bacterial swarms behave as cooperative non-equilibrium steady states (NESS). Under confinement, metabolic energy drives structured dissipative flows comprising both, convergent trapping and rotational swarming, which break detailed balance and sustain persistent motility beyond thermal limits. Noticeably, living bacterial swarms display optimized energetic performance: they cooperate to dissipate significantly less energy than the sum of uncoordinated individual contributions, achieving efficient collective motility even in confined microenvironments. This cooperative dynamics arises from adaptive flagellar coordination, regulated by structural friction and active stress, suggesting that energy-efficient swarming is a thermodynamically tuned mechanism that optimally exploits structural friction. Our findings underscore the broader role of collective motility in biology —from microbial pathogenesis to tissue dynamics, and demonstrate how stochastic thermodynamics and active field theory can reveal functional adaptations in active systems. Future work may extend this framework to eukaryote cell collectives and synthetic bioengineered systems, aiming to uncover general principles of energy allocation and control in living matter.

\section*{ Materials and Methods}
\subsection*{Bacterial samples}
We used \textit{Proteus mirabilis} due to its high swarming motility \cite{armbruster2012merging}. This rod-shaped bacterium, bearing 4–10 flagella, reaches swimming speeds up to 20~$\mu$m/s in optimal liquid media \cite{douglas1979measurement,matsuyama2000dynamic}. Cultures were grown in LB medium, refreshed biweekly, and maintained at 4$^\circ$C. For experiments, cells were incubated at 37$^\circ$C for 24 hours. Non-motile, dead bacteria were prepared by glutaraldehyde fixation (2.5\%), followed by centrifugation and resuspension in PBS \cite{rodriguez2015direct}. Experiments were performed using 20~$\mu$L of suspension in sealed chambers (1~cm$^2$ area, 1~mm thick cover glass).

\subsection*{Optical tweezers}
We employed a multi-trap optical tweezers (OT) platform (SensoCell, Impetux Optics S.L., Spain) based on photon momentum detection (PMM) \cite{farre2010force}. The setup uses a 1064 nm NIR laser (1 W at objective, 23 mW at sample), with acousto-optic beam steering and a 60x water-immersion objective (NA~1.2). Forces were measured in-plane from light momentum transfer as $\textbf{F}(x,y) = d\textbf{p}/dt$ \cite{bustamante2021optical}. At equilibrium, axial force vanishes at the trap plane ($F_z=0$). Single traps induced cluster formation at cell concentrations of 10–20 per field. The system was temperature-stabilized using a Peltier module. A CMOS camera (Thorlabs) enabled sample imaging. High-frequency force readouts (50~kHz) were digitized via a position-sensitive detector. The system was controlled through LabVIEW. PMM offers direct, calibration-free force measurements, ideal for probing bacterial motility \cite{Farré2017,jones2021stochastic}.

\subsection*{Trap stiffness}
Trap stiffness was determined via the "Particle Scan" protocol in SensoCell software \cite{farre2010force}. Forces are related to detector voltages: $\textbf{F}(x,y,t) = \hat{\alpha} V(t)$, with $\hat{\alpha}(x,y)$ a transfer function \cite{Farré2017}. Total forces decompose into static ($\textbf{F}_0 = \hat{\alpha} V_0$) and fluctuating components ($\delta \textbf{F} = \hat{\alpha} \Delta V$). Force time-series were recorded at 14~kHz for 10~s intervals. Instantaneous force is related to displacement via the Hookean trap stiffness $k$, using isotropic $\hat{\alpha}(x,y)=k$, thus $F_i = k \Delta r_i$.

\subsection*{Stochastic force correlations}
We adapted Battle et al. \cite{battle2016broken} to analyze fluctuating force trajectories in coarse-grained phase space (CGPS). Force components ($F_x$, $F_y$) were converted into probability fluxes $\textbf{J}(F_x,F_y)$ under normalized steady-state distribution $P(F_x,F_y)$. To identify dissipative features, we applied Helmholtz-Hodge decomposition: $\textbf{F} = \textbf{F}_{cons} + \textbf{F}_{diss}$ \cite{DautrayLions1990}. Curl in CGPS, $\Omega = \nabla \times \textbf{F}_{diss}$, quantified irreversibility, with flux strength approximated as $\textbf{J} \propto \Omega \Upsilon \hat{\omega}$, where $\Upsilon = \langle d\textbf{r}/d\theta \rangle$ and $\hat{\omega}$ is the angular velocity. Bootstrapping confirmed statistical robustness.

\subsection*{Optical particle tracking}
Particle tracking was performed at 65~Hz using a CCD camera (ORCA-spark, Hamamatsu), synchronized with force acquisition (14~kHz). Positions $\textbf{r}_i(t) = (x_i,y_i)$ were tracked with 100~nm resolution. Displacements were calculated as $\Delta \textbf{r}_i(t_i) = \textbf{r}_{i+1}(t_i + \Delta t) - \textbf{r}_i(t_i)$, where $\Delta t = 0.015$~s. Displacements and causative forces were synchronized via LabVIEW, enabling force-displacement correlation through $\Delta \textbf{r}_i \Leftrightarrow \textbf{F}_i/k$.

\subsection*{Stochastic energetics}
Work was computed using discrete Itô calculus as $\Delta W_i = \textbf{F}_i \circ \Delta \textbf{r}_i$, where $\circ$ denotes Stratonovich multiplication. Heat dissipation was inferred via $\tilde{Q}(t) = -\,\zeta \int_0^t \textbf{v}^2 dt'$ under Newtonian friction $\textbf{F}_{frict} = -\,\zeta \textbf{v}$. Global energy conservation $\Delta E = \tilde{W} + \tilde{Q} = 0$ allowed estimation of ensemble-averaged power dissipation: $P = \langle \Delta \,\tilde{Q}/\Delta t \rangle$ \cite{sekimoto2010stochastic}. Entropy production followed as $\sigma = \langle dQ/dt \rangle / T$ under isothermal conditions \cite{skinner2021improved}.

\subsection*{Statistical analysis}
Data are reported as mean ± SD from at least three biological replicates. Normal-distribution datasets were compared using one-way ANOVA or Student's t-test. Significance thresholds were: *$p \leq 0.05$, **$p \leq 0.01$, ***$p \leq 0.001$. All statistical analyses were performed using Prism 8.0 (GraphPad).
\bibliography{referencias}

\end{document}